\documentclass[conference,a4paper]{APSIPA2021}
\usepackage{multirow,delarray}
\usepackage{subfigure,verbatim}
\usepackage{graphicx}
\usepackage{booktabs}
\usepackage{amsmath}
\usepackage[ruled,vlined]{algorithm2e}
\usepackage{array}
\usepackage{diagbox, makecell}
\usepackage[psamsfonts]{amssymb}
\usepackage{amsxtra}
\usepackage{threeparttable}
\usepackage{colortbl}
\usepackage{lineno}
\usepackage{mathtools}
\usepackage{bm}
\usepackage{color}
\usepackage{mathrsfs}
\usepackage{algorithmic}
\usepackage{subfigure}

\usepackage{diagbox}
\usepackage{siunitx}
\usepackage{arydshln}
\usepackage{rotating}
\usepackage{booktabs}
\usepackage[misc]{ifsym}
\usepackage{bbding}
\usepackage{textcomp}
\usepackage{delarray,verbatim,booktabs}
\usepackage{lipsum}

\begin{document}

\title{CycleGAN-based Non-parallel Speech Enhancement with an Adaptive Attention-in-attention Mechanism}

\author{%
\authorblockN{%
Guochen Yu\authorrefmark{1}\authorrefmark{2}, Yutian Wang\authorrefmark{1}, Chengshi Zheng\authorrefmark{2}, Hui Wang\authorrefmark{1}, and
Qin Zhang\authorrefmark{1}
}
\authorblockA{%
\authorrefmark{1}
State Key Laboratory of Media Convergence and Communication, Communication University of China, Beijing, China \\
E-mail: \{yuguochen, wangyutian, hwang, zhangqin\}@cuc.edu.cn }
\authorblockA{%
\authorrefmark{2}
Key Laboratory of Noise and Vibration Research, Institute of Acoustics, Chinese Academy of Sciences, Beijing, China\\
E-mail: cszheng@mail.ioa.ac.cn }
}

\maketitle
\thispagestyle{empty}

\begin{abstract}
Non-parallel training is a difficult but essential task for DNN-based speech enhancement methods, for the lack of adequate noisy and paired clean speech corpus in many real scenarios. In this paper, we propose a novel adaptive attention-in-attention CycleGAN (AIA-CycleGAN) for non-parallel speech enhancement. In previous CycleGAN-based non-parallel speech enhancement methods, the limited mapping ability of the generator may cause performance degradation and insufficient feature learning. To alleviate this degradation, we propose an integration of adaptive time-frequency attention (ATFA) and adaptive hierarchical attention (AHA) to form an attention-in-attention (AIA) module for more flexible feature learning during the mapping procedure. More specifically, ATFA can capture the long-range temporal-spectral contextual information for more effective feature representations, while AHA can flexibly aggregate different AFTA's intermediate output feature maps by adaptive attention weights depending on the global context. Numerous experimental results demonstrate that the proposed approach achieves consistently more superior performance over previous GAN-based and CycleGAN-based methods in non-parallel training. Moreover, experiments in parallel training verify that the proposed AIA-CycleGAN also outperforms most advanced GAN-based and Non-GAN based speech enhancement approaches, especially in maintaining speech integrity and reducing speech distortion.
\end{abstract}

\section{Introduction}
\label{sec:intro}
Speech enhancement (SE) aims to recover clean speech components from the noise-corrupted mixture, so as to improve speech quality and intelligibility. It has become a fundamental technique in many communication applications such as the front-ends for automatic speech recognition (ASR) systems and hearing assistant devices~{\cite{loizou2013speech}}. Due to the unprecedented development of deep neural networks (DNNs), many DNN-based SE approaches have demonstrated better performance over traditional signal-processing-based approaches~{\cite{wang2018supervised}}. These DNN-based approaches can be divided into two categories, namely masking-based approaches~{\cite{wang2014training, hummersone2014ideal, hu2020dccrn}} and mapping-based approaches~{\cite{lu2013speech, xu2014regression, tan2019learning}}. Recently, Generative Adversarial Networks (GANs) have shown their promising performance in the SE area for its powerful capability of mapping the target output distribution from the original input distribution~{\cite{pascual2017segan,baby2019sergan,fu2019metricgan, liu2020cp}}, in which a generator ($G$) tries to conduct the enhancement process and a discriminator ($D$) tries to distinguish between real inputs and fake outputs generated by this generator.

\begin{figure*}[t]
	\begin{center}
		\includegraphics[width=1.8\columnwidth]{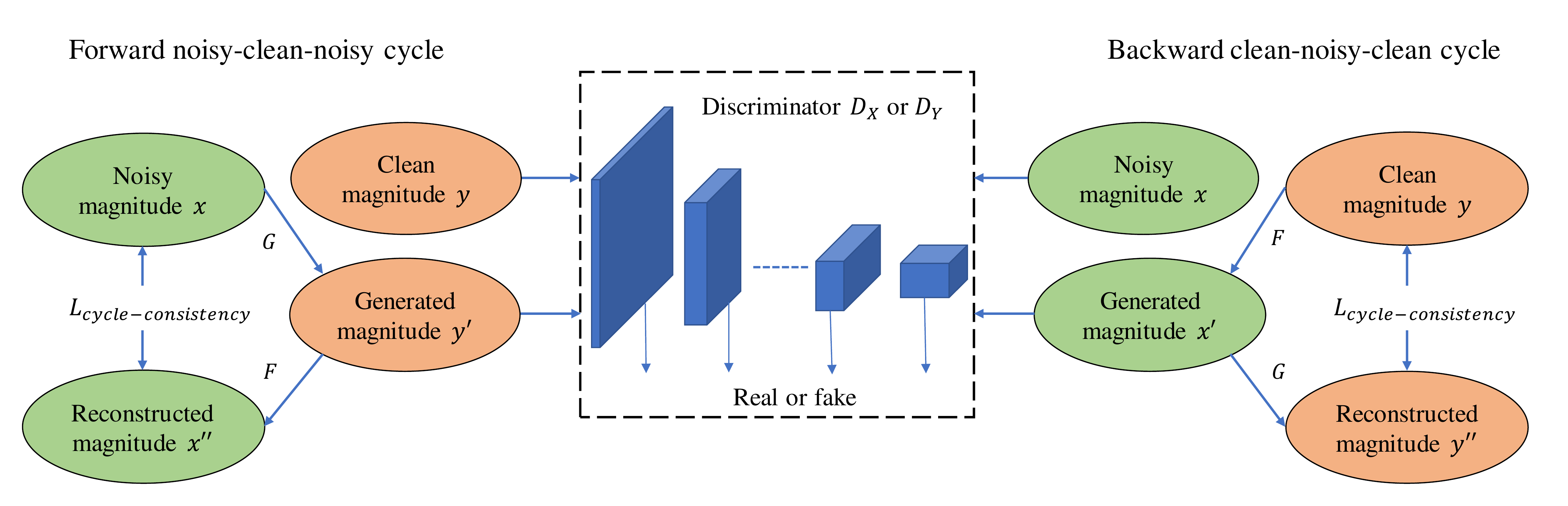}
	\end{center}

	\caption{Training procedure of the proposed method. Forward noisy-clean-noisy cycle and backward clean-noisy-clean cycle are illustrated in the left and right parts, respectively.}
	\label{fig:CycleGAN}
	
\end{figure*}
In the standard formulation of supervised speech enhancement methods, the mapping functions are trained to minimize the loss between the output features of the enhanced speech and the features of the corresponding clean speech. Therefore, they always need a large number of paired clean-noisy samples to conduct supervised training and improve the generalization of the whole network. However, there exist many practical scenarios in which it is difficult or impossible to obtain parallel recordings of clean-noisy pairs, and sometimes we can only acquire clean data that mismatches the source noisy data. To resolve this problem, CycleGAN was adopted for both standard parallel and non-parallel training in the SE area~{\cite{xiang2020parallel, yu2021two, meng2018cycle}}, which was originally proposed for unpaired image-to-image translation~{\cite{zhu2017unpaired}} and began to thrive in speech applications in recent years~{\cite{ kaneko2018CycleGAN}}. 
Nonetheless, these previous non-parallel CycleGAN-based SE methods with unpaired data can hardly achieve competitive performance when compared with the standard parallel training, because of the limited mapping ability of the generator.

In this paper, a novel CycleGAN-based system is proposed with an adaptive attention-in-attention mechanism to cope with non-parallel speech enhancement. Specifically, two generators (dubbed $G_{X\rightarrow Y}$ and $F_{ Y\rightarrow X}$) and two discriminators (dubbed $D_X$ and $D_Y$) are jointly trained with relativistic adversarial losses, cycle-consistency losses and an identity mapping loss. To improve the mapping ability of the generators, we propose a novel attention mechanism dubbed attention-in-attention (AIA) in generators for more powerful feature fusion and feature correlation learning. This AIA consists of adaptive time-frequency attention (ATFA) and adaptive hierarchical attention (AHA). Specifically, ATFA aims at capturing the long-range temporal-spectral contextual dependency in parallel, while AHA aims to flexibly aggregate all the output feature maps of ATFA together by the hierarchical attention weights depending on the global context. For discriminators, multi-scale discriminators are adopted to force the generator to pay more attention to finer details. Besides, considering the effectiveness of the power compression in the dereverberation and denoising task~{\cite{li2021importance,li2021simultaneous}}, the magnitude of the spectrum is compressed as the input features to better attenuate the background noise.

The remainder of the paper is organized as follows. In Section~{\ref{Sec2}}, the proposed framework is described in detail. The experimental setup is presented in Section~{\ref{Sec3}}, while experimental results are provided and discussed in Section~{\ref{Sec4}}. Finally, conclusions are drawn in Section~{\ref{Sec5}}.

\section{Proposed Scheme\label{Section2}}
\label{Sec2}
 \subsection{Problem Formulation}
In the speech enhancement task, when taking the short-time Fourier transform (STFT) to the mixture signal, in the time-frequency (T-F) domain, we can have:
\begin{equation}
	{X_{t,f}}=Y_{t,f}+Z_{t,f},
\end{equation}
where $X_{t,f}=\left | X_{t,f} \right |e^{j\theta_{X_{t,f}}}\in \mathbb{C}$, $Y_{t,f}=\left | Y_{t,f} \right |e^{j\theta_{Y_{t,f} }}\in \mathbb{C}$ and $Z_{t,f}=\left | Z_{t,f} \right |e^{j\theta_{Z_{t,f} }}\in \mathbb{C}$ denote the T-F representations of the mixture, clean speech and noise components in the time index of $t$ and frequency index of $f$, respectively. Most recently, using the power-compressed spectra as input features dramatically improves speech quality in the dereverberation and denoising task~{\cite{li2021importance,li2021simultaneous}}, so we conduct the power compression on the spectral magnitude before feeding into the mapping network. The compression coefficient is set to 0.5, which is reported an optimal choice in~{\cite{li2021simultaneous}. Therefore, the enhanced spectral magnitude can be expressed as:
	\begin{equation}
		\begin{gathered}	
			\left | \tilde{X}_{t,f} \right |^{\eta}=G_{X\rightarrow Y}(\left | {X}_{t,f} \right |^{\eta};\phi_G),
		\end{gathered}
	\end{equation}
	where $\eta = 0.5$; $G_{X\rightarrow Y}$ denotes the mapping function of the generator $G$ and $\phi_G$ denotes its parameter set.
	
	\begin{figure*}[!ht]
		\centering
		\centerline{\includegraphics[width=2\columnwidth]{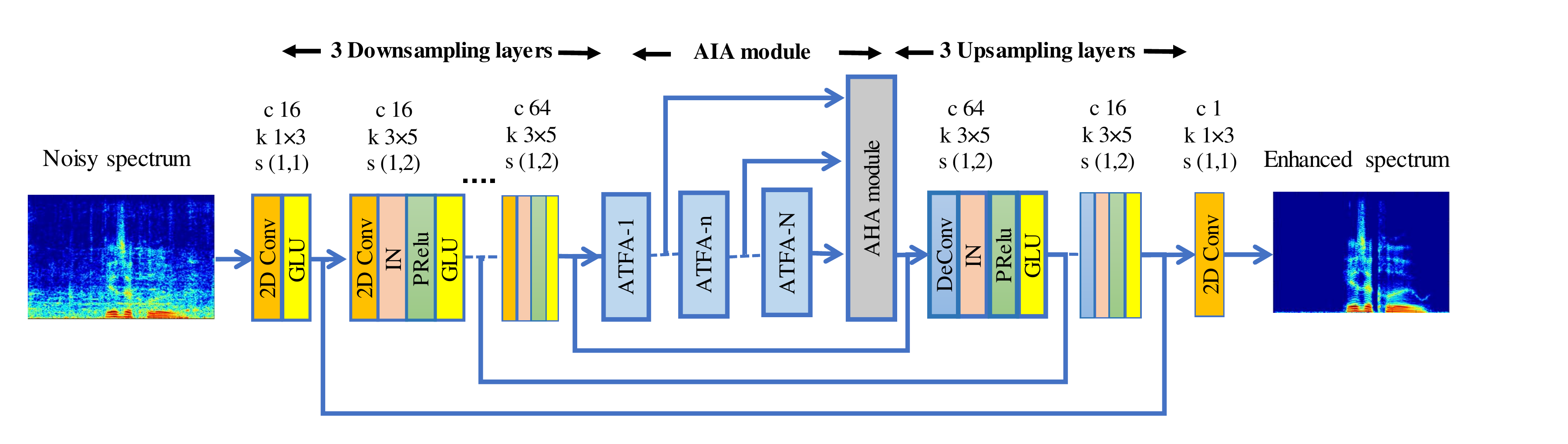}}
		\vspace{-0.3cm}
		\caption{
			The framework of the proposed generators. IN, GLU, and PRelu indicate instance normalization, gated linear unit, and parametric Relu activation, respectively.
		}
		\label{fig:generator}
	\end{figure*}
	
	\subsection{Network architecture}
	In our CycleGAN-based SE system, a forward noisy-to-clean generator $G$ is first employed to enhance the noisy features to the clean ones, while an inverse clean-to-noisy generator $F$ is applied to convert the enhanced features back to the original domain. As illustrated in Fig.~{\ref{fig:CycleGAN}}, a forward-inverse noisy-clean-noisy cycle and an inverse-forward clean-noisy-clean cycle jointly constrain $G$ and $F$ to conduct non-parallel mapping. Discriminators $D_X$ and $D_Y$ are trained to classify the target speech features as real and the generated speech features as fake. As shown in Fig.~{\ref{fig:generator}}, the generator is composed of three components, including three downsampling layers, an adaptive attention-in-attention (AIA) module and three homologous upsampling layers. Each downsampling/upsampling layer block is composed of a 2D convolution/deconvolution layer, followed by instance normalization (IN), Parametric Relu activation function (PRelu) and gated liner units (GLUs)~{\cite{dauphin2017language}}. The proposed AIA consists of six ATFA modules and an AHA module, where ATFA is proposed to capture the long-range dependencies along temporal-spectral dimensions with low computational cost and AHA is introduced to aggregate different intermediate features to capture the long-term hierarchical contextual information by the adaptive weights depending on the global context.
	
	The discriminator is composed of six 2D convolutions, each of which is followed by spectral normalization (SN) and PRelu, so as to compress the feature maps into a high-level representation. SN can stabilize the training process of the discriminator and avoid vanishing or exploding gradients~{\cite{miyato2018spectral}}. Note that we set the configuration of the utilized discriminators the same as our previous study~{\cite{wang2020improved}}. Inspired by recent studies on image enhancement~{\cite{ni2020towards}}, we propose to apply a multi-scale discriminator that uses the intermediate layer of the discriminator with a smaller receptive field, which can force the generator to produce speech features with global consistency and finer details.
	
	\begin{figure}[t]
	\centering
	\centerline{\includegraphics[width=\columnwidth]{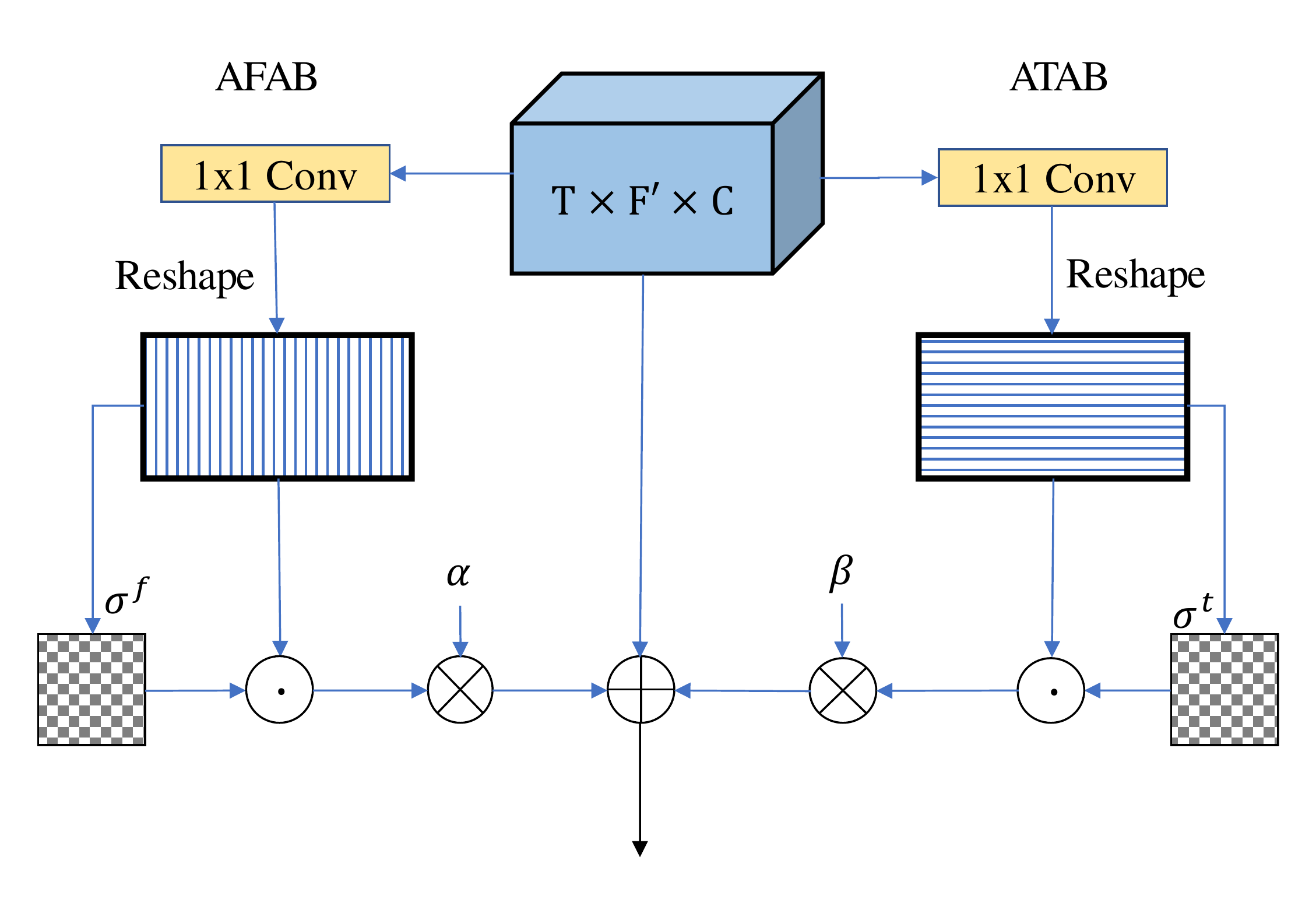}}
	\caption{The diagram of adaptive time-frequency attention (ATFA) modules. $\bigodot$ and $\bigotimes$ denote the matrix multiplication and element-wise multiplication, respectively.
	}
	\vspace{-0.3cm}
	\label{fig:ATFA}	
\end{figure}	
	\subsection{Adaptive Time-Frequency Attention}
	\label{sec:Section31}
	
	Attention mechanism~{\cite{vaswani2017attention}} has been widely used in speech processing tasks for its capability of leveraging the contextual information in the feature maps. Following the terminology in~{\cite{zhang2019self}}, we compute the attention function on the output feature maps $F_{in}\in \mathbb{R}^{B\times T\times F^{'}\times C'}$ of the downsampling layers. Here, $B$ denotes the batch size of input features, $T$ denotes the number of frames, $F^{'}$ denotes the number of frequency bins and $C'$ denotes the number of channels in each feature map. To alleviate the heavy computational complexity of conventional self-attention, we introduce an adaptive time-frequency attention (ATFA) mechanism as a lightweight solution to capture the long-range correlations exhibited in T-F spectrogram as in~{\cite{tang2020joint}}.
	As illustrated in Fig.~{\ref{fig:ATFA}}, the proposed ATFA consists of two branches: an adaptive time attention branch (ATAB) and an adaptive frequency attention branch (AFAB). The two branches cooperate to capture the global dependencies along temporal and spectral dimensions in parallel. By factorizing the original attention into time-dimension and frequency-dimension, we can reduce the large attention weight matrix to two much smaller ones, i.e., $(T\times T)$ and $(F^{'}\times F^{'})$. Along the time path, we reshape the input feature features into $C^tF'$ vectors with dimension $1 \times T$ to calculate temporal self-attention, which can be calculated as:
	\begin{equation}
		\begin{gathered}
			Q^t=Conv^{t}_Q (F_{in}), K^t=Conv^{t}_K (F_{in}), V^t=Conv^{t}_V (F_{in}),\\
			Q^t_{Res},K^t_{Res},V^t_{Res} = Reshape^t(Q^t,K^t,V^t),\\
			\sigma^{t}= softmax((Q^t_{Res})\cdot( K^t_{Res})^{T}),\\
			Out_{ATAB} = Reshape^{t'}(\sigma^{t}\cdot V^t_{Res}) ,\\
		\end{gathered}
	\end{equation}
	where $Q^t_{Res},K^t_{Res},V^t_{Res}\in R^{B\times T\times (C^t\times F')}$, $C^t=\left \{  {\frac{C}{8},\frac{C}{8},C} \right \}$, $Out_{ATAB}\in \mathbb{R}^{B\times T\times F'\times C}$, and $C=64$. Analogously, we reshape the input into $C^{f}T$ vectors with dimension $1 \times F'$ to calculate the adaptive attention $Out_{AFAB}$ along the frequency axis in parallel. Finally, The output features of these two branches and the original features are then combined together by two adaptive weights to generate the final output of ATFA module, which can be formulated as:
	\begin{equation}
		\begin{gathered}		
			Out_{ATFA} = F_{in} +\alpha Out_{ATAB} + \beta  Out_{AFAB}
		\end{gathered}
	\end{equation}
	where $F_{in}$, $Out_{ATAB}$ and $Out_{AFAB}$ represent the original input feature map given by the last downsampling layer, the output of ATAB and the output of AFAB, respectively. Here, $\alpha$ and $\beta$ are initialized as 0 and gradually lean to assign a larger weight. In summary, each branch has the following steps: (1) Reshape the input features; (2) Extract the long-range contextual dependencies along time and frequency axes, respectively; (3) Perform feature fusion along different dimensions with adaptive weights.
	
	\begin{figure*}[t]
		\centering
		\centerline{\includegraphics[width=1.6\columnwidth]{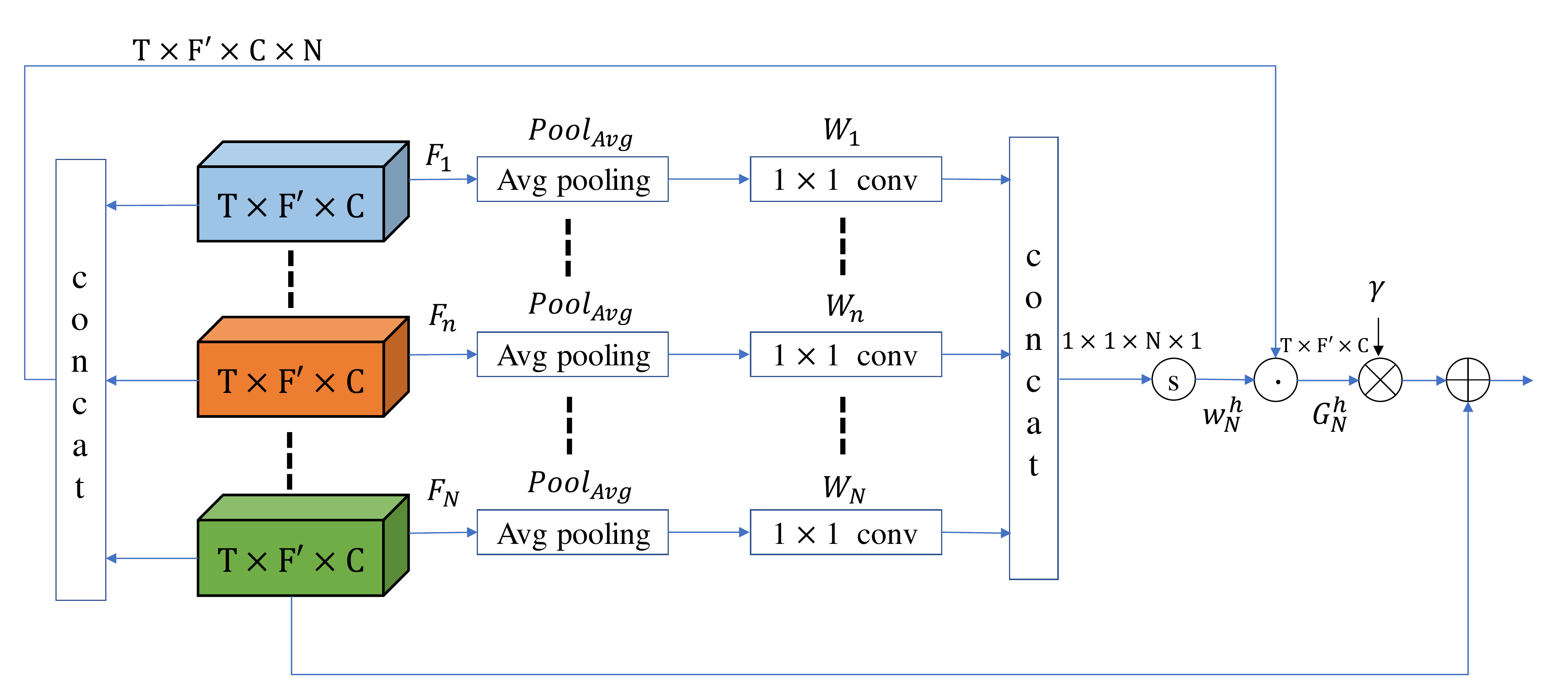}}
		\caption{The diagram of the adaptive hierarchical attention (AHA) module. $\textcircled{s}$, $\bigodot$ and $\bigotimes$ denote the softmax function, matrix multiplication and element-wise multiplication, respectively.}
		\label{fig:AHAM}
		\vspace{-0.3cm}
	\end{figure*}
	
	\subsection{Adaptive Hierarchical Attention}
	\label{sec:Section32}
	As shown in Fig.~{\ref{fig:AHAM}}, we introduce an adaptive hierarchical attention (AHA) module to integrate different hierarchical feature maps given a set of ATFA modules' outputs $F = \left \{ F_n \right \}^{N}_{n=1}, F_n \in \mathbb{R}^{T\times F'\times C}$, where $N$ is the number of ATFA blocks and set to be 6. Specifically, we first employ an average pooling layer $Pool_{Avg}$ and a $1\times1$ convolutional layer $W_{n}$ to squeeze the output feature map of each ATFA modules into a global representation: $P_n^h=Pool_{Avg}(F_n)\ast W_{n}\in R^{1\times 1\times 1}$, and then we concatenate all the outputs as $P^h \in \mathbb{R}^{1\times 1\times N\times 1}$, which is then fed into the Softmax function to obtain the hierarchical attention map $W^h \in \mathbb{R}^{1\times 1\times N\times 1}$. After that we cascade all inputs $F = \left \{ F_n \right \}^{N}_{n=1}$ to obtain a global feature map $F^h \in \mathbb{R}^{T\times F'\times C\times N}$. Subsequently, we incorporate the global contextual information by performing a matrix multiplication between $F^h \in \mathbb{R}^{T\times F'\times C\times N}$ and the hierarchical attention weights $W_N^h$, which can be defined as:
	\begin{equation}
		\begin{gathered}		
			G_{N} = \sum_{i=1}^{N}W^h_iF_i^h
		\end{gathered}
	\end{equation}
	where $G_{N}\in \mathbb{R}^{T\times F'\times C}$ denotes the global contextual feature maps. Finally, we perform an element-wise sum operation between the output feature map $F_{N}$ the last ATFA module and the global contextual feature map $G_{N}$ to obtain the final $Out_{AHA}\in \mathbb{R}^{T\times F'\times C}$:
	\begin{equation}
		\begin{gathered}		
			Out_{AHA} = F_{N} + \gamma \sum_{i=1}^{N}W^h_iF_i^h \\
			= F_{N} + \gamma \sum_{i=1}^{N} \frac{\exp(Pool_{Avg}(F_i)\ast W_i)}{\sum_{n=1}^{N}\exp(Pool_{Avg}(F_n)\ast W_n)}F_i.
		\end{gathered}
	\end{equation}
	Note that $\gamma$ is a learnable scalar coefficient and initialized as 0. This adaptive learning weight gradually learns to assign larger weight to merge global contextual information effectively. In a nutshell, $Out_{AHA}$ is a weighted sum of all ATFA modules' outputs, thus helping to fuse the global context of all intermediate feature maps with different weights and progressively guide the enhancement procedure.
	
	\subsection{Loss function}
	To ensure the effective mapping in non-parallel training, we use the following losses, namely relativistic adversarial losses, cycle-consistency losses, and an identity mapping loss, to jointly optimize the proposed model.
	
	\textbf{Relativistic adversarial loss:} For the noisy-to-clean mapping procedure, we employ the relativistic average least-square (RaLS) adversarial loss~{\cite{jolicoeur2018relativistic}} to make the enhanced compressed spectral magnitude $G_{X\rightarrow Y}(\left | {X}_{t,f} \right |^{0.5})$ appear to the target clean ones $\left | {Y}_{t,f} \right |^{0.5}$, which can be expressed as below:
	\begin{equation}
		\begin{gathered}	
			\mathcal{L}_{RA}(D_Y)= \mathbb{E}_{y}\left [(D_Y(y)-\mathbb{E}_{x}D_Y(G_{X\rightarrow Y}(x))-1)^{2}  \right ] + \\ \mathbb{E}_{x}\left [(D_Y(G_{X\rightarrow Y}(x))-\mathbb{E}_{y}(D_Y(y))+1)^{2}  \right ]
		\end{gathered}	
	\end{equation}
	\begin{equation}
		\begin{gathered}	
			\mathcal{L}_{RA}(G_{X\rightarrow Y})=  \mathbb{E}_{x}\left [(D_Y(G_{X\rightarrow Y}(x)))-\mathbb{E}_{y}D_Y(y) -1)^{2} \right] \\+ \mathbb{E}_{y}\left [(D_Y(y)-\mathbb{E}_{x}(D_Y(G_{X\rightarrow Y}(x)))+1)^{2}  \right ]
		\end{gathered}
	\end{equation}	
where $x$ and $y$ are the compressed magnitudes of noisy and clean spectrum (i.e. $\left | {X}_{t,f} \right |^{0.5}$ and $\left | {Y}_{t,f} \right |^{0.5}$), respectively. Here, the generator $G_{X\rightarrow Y}$ tries to synthesize the enhanced spectral magnitude that can deceive the discriminator $D_Y$ by minimizing $\mathcal{L}_{RA}(G_{X\rightarrow Y})$, whereas the discriminator $D_Y$ attempts to distinguish the generated spectral magnitude from the clean one $y$ by minimizing $\mathcal{L}_{RA}(D_Y)$. Analogously, the inverse clean-to-noisy generator $F_{ Y\rightarrow X}$ and its corresponding discriminator $D_X$ are optimized using $\mathcal{L}_{RA}(D_X)$ and $\mathcal{L}_{RA}(F_{ Y\rightarrow X})$.
	
\textbf{Cycle-consistency loss:} Without parallel supervision, generators may map source feature space to any random permutation of the target space within only adversarial losses. To constrain the non-parallel mapping, a cycle-consistency loss is utilized to bring the output back to original input data. The cycle-consistency loss $\mathcal{L}_{cycle}(G_{X\rightarrow Y},F_{ Y\rightarrow X})$ can help two generators $G_{X\rightarrow Y}$ and $F_{ Y\rightarrow X}$ to identity the pseudo pair $(x,y)$ without paired data as follows:	
	\begin{equation}
		\begin{gathered}
			\mathcal{L}_{cycle}(G_{X\rightarrow Y},F_{ Y\rightarrow X})=\mathbb{E}_{x}\left [\left \|F_{ Y\rightarrow X}(G_{X\rightarrow Y}(x)) -x \right \|_{1}\right ] \\+\mathbb{E}_{y}\left [\left \|G_{X\rightarrow Y}(F_{ Y\rightarrow X}(y)) -y \right \|_{1}\right ]
		\end{gathered}
	\end{equation}
	where $\left \| \cdot  \right \|_1$ indicates the $L1$ Norm.
	
	\textbf{Identity-mapping loss:} Since the generator should not modify the compositions, such as linguistic information of the target speech feature when it is fed into the generator as the input~{\cite{meng2018cycle}}, too much, we regularize generators $G$ and $F$ to be as close as possible to the identity mapping by minimizing an identity-mapping loss~{\cite{zhu2017unpaired}}, which can be given by:
	\begin{equation}
		\begin{gathered}
			\mathcal{L}_{id}(G_{X\rightarrow Y},F_{ Y\rightarrow X})=\mathbb{E}_{x}\left [\left \|F_{ Y\rightarrow X}(x) -x \right \|_{1}\right ]+ \\ \mathbb{E}_{y}\left [\left \|G_{X\rightarrow Y}(y) -y \right \|_{1}\right ]
		\end{gathered}
	\end{equation}
	where the magnitudes of the target spectrum (i.e., $y$ and $x$) are provided as the inputs of the generators (i.e., $G_{X\rightarrow Y}$ and $F_{ Y\rightarrow X}$), respectively. In summary, the overall loss can be summarized as follows:
	\begin{equation}
		\begin{gathered}	
			\mathcal{L}_{Full}=\mathcal{L}_{RA}(G_{X\rightarrow Y},D_Y) + \mathcal{L}_{RA}(F_{ Y\rightarrow X},D_X) \\ +\lambda _{cycle}\mathcal{L}_{cycle}(G_{X\rightarrow Y},F_{ Y\rightarrow X}) +\lambda _{id}\mathcal{L}_{id}(G_{X\rightarrow Y},F_{ Y\rightarrow X})
		\end{gathered}
	\end{equation}
	where $\lambda_{cycle}$ and $\lambda_{id}$ are tunable hyper-parameters, which are set to be 5 and 10, respectively.

\section{Experiments\label{Section3}}
\label{Sec3}
\vspace{-0.3cm}
\subsection{Datasets\label{Section31}}
 The dataset used in this work is publicly available as proposed in~{\cite{valentini2016investigating}}, which is a selection of the Voice Bank corpus~{\cite{veaux2013voice}} with 28 speakers for training and another 2 unseen speakers for the test. The training set consists of 11572 mono audio samples, while the test set contains 2 speakers' (one male and one female) 824 utterances. For the training set, audio samples are mixed together with one of the 10 noise types, i.e., two artificial (babble and speech shaped) and eight real noise from the DEMAND database~{\cite{thiemann2013diverse}}, at four SNRs of 0, 5, 10 and 15 dB. The test utterances are created with 5 unseen test-noise types (all from the DEMAND database) at SNRs of 2.5, 7.5, 12.5 and 17.5 dB. The original raw waveforms are downsampled from 48kHz to 16kHz beforehand.

\renewcommand\arraystretch{1}
\begin{table}[t!]
	\setcounter{table}{0}
	\caption{Ablation study for Normal and Compressed magnitude under non-parallel training.}
	\centering
	\scalebox{0.95}{
		\begin{tabular}{cccccc}
			\toprule
			Models
			&PESQ &SSNR  &STOI & DNSMOS\\
			\midrule
			Unprocessed  &1.97 &1.68 &0.921 &3.02\\
			\midrule
			\multicolumn{6}{c}{\textbf{Normal magnitude}} \\
			\midrule
			CycleGAN (baseline)  &2.47 &5.69 &0.924 &3.36\\
			\midrule
			CycleGAN+ATAB  &2.53 &6.28 &0.929 &3.40 \\
			\midrule
			CycleGAN+AFAB  &2.54 &6.42 &0.928 &3.39 \\
			\midrule
			CycleGAN+ATFA   &2.59 &6.65 &\textbf{0.932} &3.41 \\ 			
			\midrule
			AIA-CycleGAN  &2.61 &6.69 &0.931 &3.43 \\
			
			\midrule
			\multicolumn{6}{c}{\textbf{Compressed magnitude (parameter $\eta =0.5$)}} \\
			\midrule
			CycleGAN (baseline)&2.56 &6.21 &0.926 &3.38\\
			\midrule
			CycleGAN+ATAB  &2.59 &6.67 &0.928  &3.41\\
			\midrule
			CycleGAN+AFAB   &2.61 &6.61 &0.931  &3.43\\
			\midrule
			CycleGAN+ATFA  &2.64 &6.94 &0.930 &3.45 \\ 			
			\midrule
			AIA-CycleGAN  &\textbf{2.67} &\textbf{7.23} &\textbf{0.932} &\textbf{3.47} \\
			\bottomrule
	\end{tabular}}
	\label{tbl:ablation-study1}
	\vspace{-0.5cm}
\end{table}

\vspace{-0.1cm}
\subsection{Implementation Setup\label{Section42}}
The Hanning window of length 32ms is utilized, with 75\% overlap between adjacent frames. The 512-point STFT is utilized and the 257-dimension compressed spectral magnitude is used as the input feature. For the non-parallel training strategy, we randomly crop a fixed-length segment (i.e., 108 frames) from a randomly selected noisy audio file as the input, while the target is a randomly selected clean audio file that is different from the input audio. That is to say, the input totally mismatches the target speech features. We adopt the Adam optimizer~{\cite{kingma2014adam}} with the momentum term $\beta_{1}=0.9$, $\beta_{2}=0.999$ and train the networks with an initial learning rate of 0.0001 for discriminators and 0.0002 for generators, respectively. The same learning rates are maintained for the first 50 epochs, while they linearly decay in the remaining iterations. We set the batch size to 4 and use $\mathcal{L}_{id}$ only for the first 20 epochs.

\vspace{-0.2cm}
\section{Results and Analysis\label{Section4}}
\label{Sec4}
We use the following objective metrics to evaluate speech enhancement performance: the perceptual evaluation of speech quality (PESQ)~{\cite{rix2001perceptual}}, short-time objective intelligibility (STOI)~{\cite{taal2010short}}, segmental signal-to-noise ratio (SSNR), the mean opinion score (MOS) prediction of the speech signal distortion (CSIG)~{\cite{hu2007evaluation}}, the MOS prediction of the intrusiveness of background noise (CBAK) and the MOS prediction of the overall effect (COVL)~{\cite{hu2007evaluation}}. In addition, we evaluate the subjective quality by DNSMOS~{\cite{reddy2020dnsmos}}, which is a robust non-intrusive perceptual speech quality metric designed to evaluate noise suppressors. Higher values of all metrics indicate better performance.

\subsection{Ablation study\label{Section41}}
We first investigate the effectiveness of the proposed attention modules and the power compression. As shown in Table~{\ref{tbl:ablation-study1}}, we set the CycleGAN without proposed attention modules as the baseline, which is also trained with the relativistic average least-square loss. From the results, we can have the following observations. Firstly, compared with non-compressed methods, CycleGAN-based approaches fed with the compressed spectral magnitude achieve better performance, indicating that the power compression facilitates more accurate spectrum recovery. For example, compressed-magnitude CycleGAN achieves average 0.09 PESQ and 0.52dB SSNR improvements over the normal-magnitude CycleGAN. The possible rationale is that, when the compression operation is applied, the gap between the magnitude of the speech and noise components is narrowed, that is to say, the residual noise components in the weak energy regions (i.e., middle and high-frequency regions) are given more priority during the enhancement produce~{\cite{li2021importance}}. Secondly, by adding the proposed attention branches, CycleGAN+ATAB improves the average PESQ and STOI by 0.03 and 0.002 over the compressed-magnitude baseline, while CycleGAN + AFAB improves the average PESQ and STOI by 0.05 and 0.004. This indicates that the ATAB and AFAB can effectively guide the feature learning procedure of the generators. By integrating ATAB and ATFB as the ATFA modules, we also observe considerable improvements on PESQ, SSNR and DNSMOS scores. Finally, by integrating the AHA module and ATFA modules as an AIA module, we can see that the proposed AIA-CycleGAN significantly outperforms other comparisons, which verifies the effectiveness of the proposed AIA module in improving the speech quality.

\vspace{-0.1cm}
\subsection{Comparison under non-parallel and parallel training\label{Section53}}
\vspace{-0.1cm}
To investigate the effectiveness of our proposed method with both the parallel and non-parallel training strategy, we compare our proposed method with the reference methods including conventional GANs and CycleGANs. Here, "GAN-normal" is fed with the normal magnitude as the input features while "GAN-compressed" is fed with the compressed magnitude. From Table~{\ref{results}}, we can observe the following two phenomena. Firstly, GAN-based methods yield similar performance with CycleGAN in the parallel training, whereas CycleGAN outperforms GAN-based methods by a large margin in the non-parallel training. This indicates the cycle consistency constraint can prevent the mapping ability of the generators from sharply degrading under unpaired data. Besides, by employing the AIA module, the proposed approach under unpaired data achieves similar and competitive performance compared with parallel training, and significantly outperforms all the baselines. For example, AIA-CycleGAN surpasses GAN-compressed by a large margin in PESQ, CSIG, CBAK, COVL and DNSMOS using non-parallel training, which is 0.64, 0.38, 0.46, 0.54 and 0.79, respectively.

Fig.~{\ref{fig:p232_010_CycleGAN}} shows the spectrograms of the noisy/clean utterances and the utterances enhanced by GAN-compressed, CycleGAN and AIA-CycleGAN in the non-parallel training. This figure demonstrates that CycleGANs dramatically surpass the conventional GAN-based method. Moreover, we observe that AIA-CycleGAN can effectively surpass the original CycleGAN in terms of noise suppression. For example, as shown in the red sign area and green sign area of Fig.~{\ref{fig:p232_010_CycleGAN}} (c) and (d), AIA-CycleGAN shows a more powerful capability of suppressing residual noise components.

\renewcommand\arraystretch{1.2}
\begin{table}[t!]
	\caption{Experimental results among different models under non-parallel and parallel training. \textbf{bold} indicates the best results for different training conditions. }
	\label{results}
	\centering
	\small
	\scalebox{0.8}{
		\begin{tabular}{lcccccc}
			\hline
			\multicolumn{1}{l|}{Methods} & PESQ & STOI  & CSIG & CBAK & COVL &DNSMOS\\ \hline
			\multicolumn{1}{l|}{Noisy} & 1.97 & 0.921 & 3.35 & 2.44 & 2.63 & 3.02\\ \hline
			\multicolumn{7}{c}{\textbf{Non-parallel training}} \\ \hline
			\multicolumn{1}{l|}{GAN-Normal} & 2.01 & 0.914 & 3.48 & 2.74 & 2.67 & 2.68\\ \hline
			\multicolumn{1}{l|}{GAN-Compressed } &2.03 & 0.916 & 3.54 & 2.78 & 2.72 & 2.72\\ \hline
			\multicolumn{1}{l|}{CycleGAN} & 2.56 & 0.927 & 3.78 & 3.14 & 3.16 & 3.38 \\ \hline
			\multicolumn{1}{l|}{AIA-CycleGAN } & \textbf{2.67} & \textbf{0.932} & \textbf{3.86} & \textbf{3.20} & \textbf{3.21} & \textbf{3.47}\\ \hline 		
			\multicolumn{7}{c}{\textbf{Parallel training}} \\ \hline			
			\multicolumn{1}{l|}{GAN-Normal} & 2.56 & 0.931 & 3.72 & 3.23 & 3.16 & 3.33 \\  \hline
			\multicolumn{1}{l|}{GAN-Compressed } & 2.60 & 0.934 & 3.78 & \textbf{3.25} & 3.18 & 3.35 \\ \hline
			\multicolumn{1}{l|}{CycleGAN} & 2.62 & 0.932 & 3.87 & 3.16 & 3.24 & 3.41\\ \hline
			\multicolumn{1}{l|}{AIA-CycleGAN} &\textbf{2.74} & \textbf{0.936} & \textbf{3.96} & \textbf{3.25} & \textbf{3.29} & \textbf{3.49}\\ \hline
		\end{tabular}
	}
\vspace{-0.4cm}
\end{table}

\begin{figure}
	\centering
	\centerline{\includegraphics[width=\columnwidth]{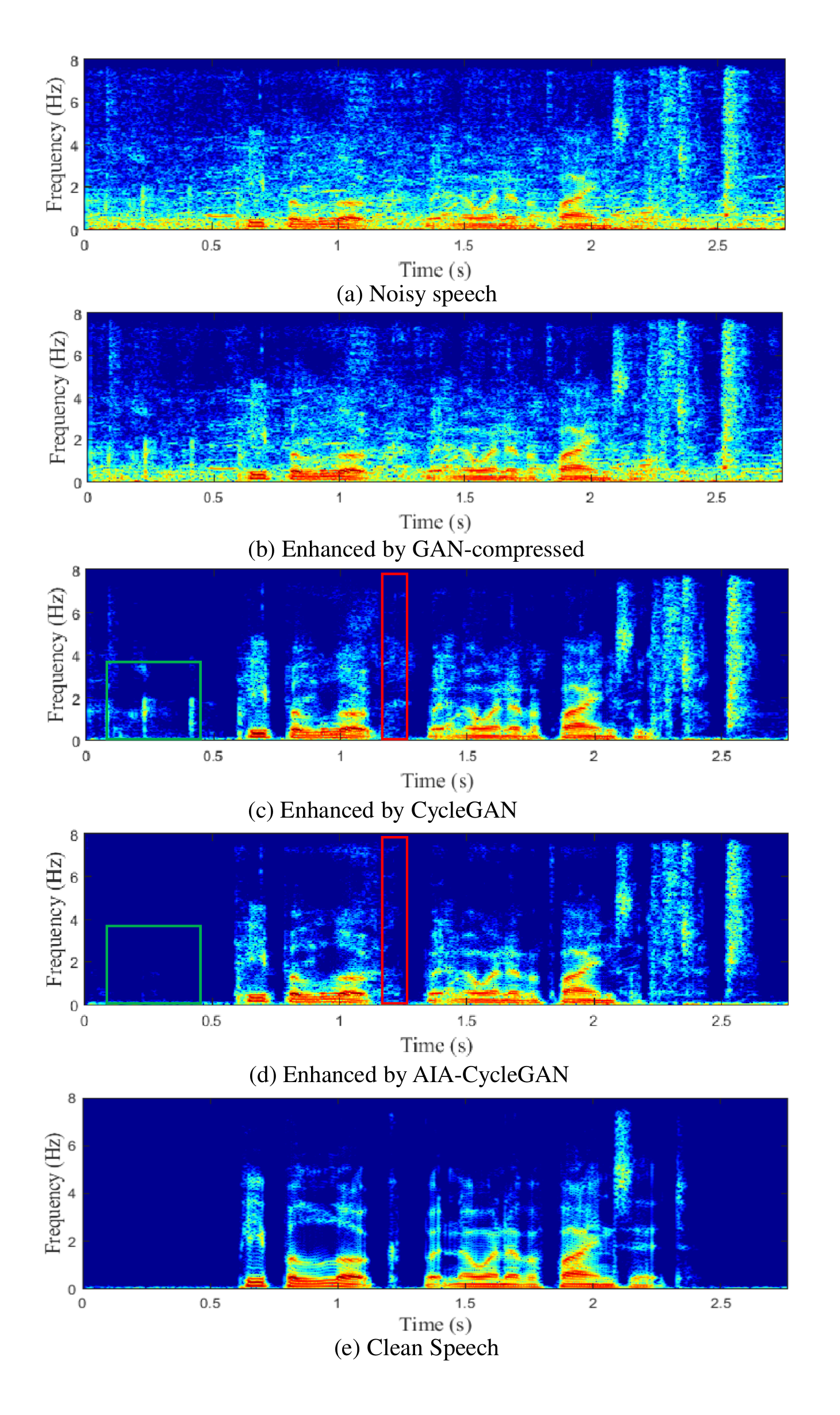}}
	\vspace{-0.6cm}
	\caption{Visualization of the noisy/clean spectrogram and enhanced spectrogram using different methods.  }
	\label{fig:p232_010_CycleGAN}
	\vspace{-0.3cm}
\end{figure}

\vspace{-0.1cm}
\subsection{Comparison with other competitive GAN-based and Non-GAN based approaches\label{Section54}}
\vspace{-0.1cm}
Our proposed model is also compared with several other competitive GAN-based and Non-GAN based baselines under standard paired data. As seen from Table~{\ref{tbl:VB-results}}, AIA-CycleGAN outperforms these advanced GAN-based methods in terms of PESQ, CSIG and COVL by considerable improvements, while providing similar STOI and CBAK with CP-GAN~{\cite{liu2020cp}}. For example, AIA-CycleGAN provides average 0.58 PESQ, 0.011 STOI, 0.48 CSIG, 0.31 CBAK and 0.49 COVL improvements than SEGAN~{\cite{pascual2017segan}}, which is the first GAN-based SE approach in the time domain. The significant improvements in PESQ, CSIG and COVL indicate that our proposed method better maintains speech integrity while reducing speech distortion. When compared with other Non-GAN based methods, our AIA-CycleGAN also provides competitive performance especially in terms of PESQ and CSIG scores. Note that we reimplement GCRN~{\cite{tan2019learning}} and DCCRN~{\cite{hu2020dccrn}} in Voice bank + DEMAND dataset, and directly use the reported scores of other methods in their original papers.

\renewcommand\arraystretch{1.2}
\begin{table}[t!]
	\caption{Experimental results among different models under parallel training. }
	\vspace{-3mm}

	\label{tbl:VB-results}
	\centering
	\small
	\scalebox{0.9}{
		\begin{tabular}{lcccccc}
			\hline
			\multicolumn{1}{l|}{Methods} & PESQ & STOI  & CSIG & CBAK & COVL \\ \hline
			\multicolumn{1}{l|}{Noisy}  & 1.97 & 0.921 & 3.35 & 2.44 & 2.63 \\ \hline
			\multicolumn{6}{c}{\textbf{GAN-based methods}} \\ \hline
			\multicolumn{1}{l|}{SEGAN~{\cite{pascual2017segan}}}  & 2.16 & 0.925 & 3.48 & 2.94 & 2.80 \\
			\multicolumn{1}{l|}{MMSEGAN~{\cite{soni2018time}} } & 2.53 & 0.930 & 3.80 & 3.12 & 3.14 \\
			\multicolumn{1}{l|}{RSGAN~{\cite{baby2019sergan}}}   & 2.51 & 0.937 & 3.78 & 3.23 & 3.16 \\
			\multicolumn{1}{l|}{RaSGAN~{\cite{baby2019sergan}} } & 2.57 & 0.937 & 3.83 & 3.28 & 3.20 \\
			\multicolumn{1}{l|}{CP-GAN~{\cite{liu2020cp}} }  & 2.64 & \textbf{0.940} & 3.93 & 3.29 & 3.28 \\ \hline
			\multicolumn{6}{c}{\textbf{Non-GAN based methods}} \\ \hline			\multicolumn{1}{l|}{Wave-U-net~{\cite{stoller2018wave}} }  & 2.64 & \makecell[c]{--} & 3.56 & 3.08 & 3.09 \\  
			\multicolumn{1}{l|}{DFL-SE~{\cite{germain2019speech}} }  &\makecell[c]{--} & \makecell[c]{--} & 3.86 & \textbf{3.33} & 3.22 \\ 
			\multicolumn{1}{l|}{CRN-MSE~{\cite{tan2018convolutional}} }  & 2.61 & 0.938 & 3.78 & 3.11 & 3.24 \\ 
			\multicolumn{1}{l|}{GCRN~{\cite{tan2019learning}} }  & 2.51  & \textbf{0.940} & 3.71 & 3.24 & 3.09 \\ 			
			\multicolumn{1}{l|}{DCCRN~{\cite{hu2020dccrn}} } & 2.68 & 0.939 & 3.88 & 3.18 & 3.27 \\ \hline			
			\multicolumn{1}{l|}{AIA-CycleGAN} & \textbf{2.74} & 0.936 & \textbf{3.96} & 3.25 & \textbf{3.29} \\ \hline
		\end{tabular}
	}
	\vspace{-5mm}
\end{table}

\section{Conclusions\label{Section5}}
\label{Sec5}
\vspace{-1mm}
In this paper, we propose a novel adaptive attention-in-attention CycleGAN (AIA-CycleGAN) to solve the difficulty in non-parallel speech enhancement task. Specifically, we use relativistic adversarial losses, cycle-consistency losses and an identity loss to jointly constrain the forward noisy-clean-noisy cycle and backward clean-noisy-clean cycle. To effectively improve the feature correlation learning in the generators, we integrate adaptive time-frequency attention and adaptive hierarchical attention to form an attention-in-attention module to capture local and global long-range dependencies. By employing ATFA, generators can capture the long-range temporal and frequency contextual information to distinguish different types of information for more effective feature representations. By employing AHA, generators can capture the long-range hierarchical contextual information to flexibly aggregate different global feature maps by learnable weights. Experimental results demonstrate that the proposed approach provides consistently better speech enhancement performance than the previous GAN-based and CycleGAN-based baselines under both parallel and non-parallel training.

\vspace{-1mm}
\section*{Acknowledgment}
This work was supported in part by the National Natural Science Foundation of China under Grant 61631016 and Grant 61501410, and in part by the Fundamental Research Funds for the Central Universities under Grant 3132018XNG1805.


\vspace{-3mm}

\begin{thebibliography}{1}
	
\bibitem{loizou2013speech}
P.~C. Loizou,
\newblock {\em Speech enhancement: theory and practice},
\newblock CRC press, 2013.

\bibitem{wang2018supervised}
D.~L. Wang and J.~Chen,
\newblock ``Supervised speech separation based on deep learning: An overview,''
\newblock {\em  IEEE/ACM Trans. Audio. Speech, Lang. Process.}, vol. 26, no. 10, pp.
1702--1726, 2018.

\bibitem{wang2014training}
Y. Wang, A. Narayanan, and D. L. Wang,
\newblock ``On training targets for supervised speech separation,''
\newblock {\em  IEEE/ACM Trans. Audio. Speech, Lang. Process.}, vol. 22, no. 12, pp. 1849--1858, 2014.

\bibitem{hummersone2014ideal}
C.~Hummersone, T.~Stokes, and T.~Brookes,
\newblock ``On the ideal ratio mask as the goal of computational auditory scene
analysis,''
\newblock in {\em Blind source separation}, pp. 349--368. Springer, 2014.

\bibitem{hu2020dccrn}
Y.~Hu, Y.~Liu, S.~Lv, M.~Xing, and L.~Xie,
\newblock ``Dccrn: Deep complex convolution recurrent network for phase-aware
speech enhancement,''
\newblock {\em arXiv preprint arXiv:2008.00264}, 2020.	

\bibitem{lu2013speech}
X. Lu, Y. Tsao, S. Matsuda, and C. Hori,
\newblock ``Speech enhancement based on deep denoising autoencoder.,''
\newblock in {\em Proc. of Interspeech}, 2013, vol. 2013, pp. 436--440.

\bibitem{xu2014regression}
Y. Xu, J. Du, L. R. Dai, and C. H. Lee,
\newblock ``A regression approach to speech enhancement based on deep neural
networks,''
\newblock {\em  IEEE/ACM Trans. Audio. Speech, Lang. Process.}, vol. 23, no. 1, pp. 7--19, 2014.

\bibitem{tan2019learning}
K.~Tan and D.~L. Wang,
\newblock ``Learning complex spectral mapping with gated convolutional
recurrent networks for monaural speech enhancement,''
\newblock {\em  IEEE/ACM Trans. Audio. Speech, Lang. Process.}, vol. 28, pp. 380--390, 2019.

\bibitem{pascual2017segan}
S. Pascual, A. Bonafonte, and J. Serra,
\newblock ``Segan: Speech enhancement generative adversarial network,''
\newblock {\em Proc. of Interspeech }, pp. 3642--3646, 2017.

\bibitem{baby2019sergan}
D. Baby and S. Verhulst,
\newblock ``Sergan: Speech enhancement using relativistic generative
adversarial networks with gradient penalty,''
\newblock in {\em Proc. of ICASSP}. IEEE, 2019, pp. 106--110.

\bibitem{fu2019metricgan}
S. W. Fu, C. F. Liao, Y.~Tsao, and S. D. Lin,
\newblock ``Metricgan: Generative adversarial networks based black-box metric
scores optimization for speech enhancement,''
\newblock in {\em Proc. IMCL}. PMLR, 2019,
pp. 2031--2041.

\bibitem{liu2020cp}
G. Liu, K.~Gong, X. Liang, and Z. Chen,
\newblock ``Cp-gan: Context pyramid generative adversarial network for speech
enhancement,''
\newblock in {\em Proc. of ICASSP}. IEEE, 2020, pp. 6624--6628.

\bibitem{soni2018time}
M.~H. Soni, N. Shah, and H.~A. Patil,
\newblock ``Time-frequency masking-based speech enhancement using generative
adversarial network,''
\newblock in {\em Proc. of ICASSP}. IEEE, 2018, pp. 5039--5043.

\bibitem{xiang2020parallel}
Y. Xiang and C. Bao,
\newblock ``A parallel-data-free speech enhancement method using
multi-objective learning cycle-consistent generative adversarial network,''
\newblock {\em  IEEE/ACM Trans. Audio. Speech, Lang. Process.}, vol. 28, pp. 1826--1838, 2020.

\bibitem{yu2021two}
G.~Yu, Y.~Wang, H.~Wang, Q.~Zhang, and C.~Zheng,
\newblock ``A two-stage complex network using cycle-consistent generative
adversarial networks for speech enhancement,''
\newblock {\em arXiv preprint arXiv:2109.02011}, 2021.

\bibitem{meng2018cycle}
Z. Meng, J. Li, Y. Gong, et~al.,
\newblock ``Cycle-consistent speech enhancement,''
\newblock {\em arXiv preprint arXiv:1809.02253}, 2018.

\bibitem{zhu2017unpaired}
J. Y. Zhu, T. Park, P. Isola, and A.~A. Efros,
\newblock ``Unpaired image-to-image translation using cycle-consistent
adversarial networks,''
\newblock in {\em Proc. of ICCV}, 2017, pp. 2223--2232.


\bibitem{kaneko2018CycleGAN}
T. Kaneko and H. Kameoka,
\newblock ``CycleGAN-vc: Non-parallel voice conversion using cycle-consistent
adversarial networks,''
\newblock in {\em 2018 26th European Signal Processing Conference (EUSIPCO)}.
IEEE, 2018, pp. 2100--2104.


\bibitem{li2021importance}
A. Li, C. Zheng, R. Peng, and X. Li,
\newblock ``On the importance of power compression and phase estimation in
monaural speech dereverberation,''
\newblock {\em JASA Express Letters}, vol. 1, no. 1, pp. 014802, 2021.

\bibitem{li2021simultaneous}
A. Li, W. Liu, X. Luo, G. Yu, C. Zheng, and X. Li,
\newblock ``A simultaneous denoising and desreverberation framework with target
decoupling,''
\newblock {\em arXiv preprint arXiv:2106.12743}, 2021.

\bibitem{dauphin2017language}
Y.~N. Dauphin, A. Fan, M. Auli, and D. Grangier,
\newblock ``Language modeling with gated convolutional networks,''
\newblock in {\em Proc. IMCL}. PMLR, 2017,
pp. 933--941.

\bibitem{miyato2018spectral}
T. Miyato, T. Kataoka, M. Koyama, and Y. Yoshida,
\newblock ``Spectral normalization for generative adversarial networks,''
\newblock in {\em Proc. ICLR}, 2018.

\bibitem{wang2020improved}
Y. Wang, G. Yu, J. Wang, H. Wang, and Q. Zhang,
\newblock ``Improved relativistic cycle-consistent gan with dilated residual
network and multi-attention for speech enhancement,''
\newblock {\em IEEE Access}, vol. 8, pp. 183272--183285, 2020.

\bibitem{ni2020towards}
Z. Ni, W. Yang, S. Wang, L. Ma, and S. Kwong,
\newblock ``Towards unsupervised deep image enhancement with generative
adversarial network,''
\newblock {\em IEEE Transactions on Image Processing}, vol. 29, pp. 9140--9151,
2020.

\bibitem{vaswani2017attention}
A. Vaswani, N. Shazeer, N. Parmar, J. Uszkoreit, L. Jones,
A.~N. Gomez, L. Kaiser, and I. Polosukhin,
\newblock ``Attention is all you need,''
\newblock {\em arXiv preprint arXiv:1706.03762}, 2017.

\bibitem{zhang2019self}
H. Zhang, I. Goodfellow, D. Metaxas, and A. Odena,
\newblock ``Self-attention generative adversarial networks,''
\newblock in {\em Proc. IMCL}. PMLR, 2019,
pp. 7354--7363.

\bibitem{tang2020joint}
C. Tang, C. Luo, Z. Zhao, W. Xie, and W. Zeng,
\newblock ``Joint time-frequency and time domain learning for speech
enhancement,''
\newblock in {\em Proc. of AAAI}, 2020, pp. 3816--3822.

\bibitem{jolicoeur2018relativistic}
A. Jolicoeur-Martineau,
\newblock ``The relativistic discriminator: a key element missing from standard
gan,''
\newblock {\em arXiv preprint arXiv:1807.00734}, 2018.

\bibitem{valentini2016investigating}
C. Valentini-Botinhao, X. Wang, S. Takaki, and J. Yamagishi,
\newblock ``Investigating rnn-based speech enhancement methods for noise-robust
text-to-speech.,''
\newblock in {\em Proc. SSW}, 2016, pp. 146--152.

\bibitem{veaux2013voice}
C.~Veaux, J.~Yamagishi, and S.~King,
\newblock ``The voice bank corpus: Design, collection and data analysis of a
large regional accent speech database,''
\newblock in {\em Proc. O-COCOSDA/CASLRE}. IEEE, 2013, pp. 1--4.

\bibitem{thiemann2013diverse}
J. Thiemann, N. Ito, and E. Vincent,
\newblock ``The diverse environments multi-channel acoustic noise database: A
database of multichannel environmental noise recordings,''
\newblock {\em Acoustical Society of America Journal}, vol. 133, no. 5, pp.
3591, 2013.

\bibitem{kingma2014adam}
D.~Kingma and J.~Ba,
\newblock ``Adam: A method for stochastic optimization,''
\newblock {\em arXiv preprint arXiv:1412.6980}, 2014.

\bibitem{rix2001perceptual}
A.~Rix, J.~Beerends, M.~Hollier, and A.~Hekstra,
\newblock ``Perceptual evaluation of speech quality ({PESQ})-a new method for
speech quality assessment of telephone networks and codecs,''
\newblock in {\em Proc. of ICASSP}. IEEE, 2001, vol.~2, pp. 749--752.

\bibitem{taal2010short}
C.~H. Taal, R.~C. Hendriks, R. Heusdens, and J. Jensen,
\newblock ``A short-time objective intelligibility measure for time-frequency
weighted noisy speech,''
\newblock in {\em Proc. of ICASSP}. IEEE, 2010, pp. 4214--4217.

\bibitem{hu2007evaluation}
Y.~Hu and P.~C. Loizou,
\newblock ``Evaluation of objective quality measures for speech enhancement,''
\newblock {\em  IEEE/ACM Trans. Audio. Speech, Lang. Process.},
vol. 16, no. 1, pp. 229--238, 2007.

\bibitem{reddy2020dnsmos}
C.~K. Reddy, V. Gopal, and R. Cutler,
\newblock ``Dnsmos: A non-intrusive perceptual objective speech quality metric
to evaluate noise suppressors,''
\newblock {\em arXiv preprint arXiv:2010.15258}, 2020.


\bibitem{stoller2018wave}
D.~Stoller, S.~Ewert, and S.~Dixon,
\newblock ``Wave-u-net: A multi-scale neural network for end-to-end audio
source separation,''
\newblock {\em arXiv preprint arXiv:1806.03185}, 2018.


\bibitem{germain2019speech}
F.~Germain, Q.~Chen, and V.~Koltun,
\newblock ``Speech denoising with deep feature losses,''
\newblock in {\em Proc. of Interspeech}, 2019, pp. 2723--2727.


\bibitem{tan2018convolutional}
K.~Tan and D.~L. Wang,
\newblock ``A convolutional recurrent neural network for real-time speech
enhancement.,''
\newblock in {\em Proc. of Interspeech}, 2018, pp. 3229--3233.


\end{thebibliography}
\end{document}